\documentclass[preprint,onecolumn,nofootinbib]{revtex4}
\pdfoutput=1
\usepackage[colorlinks=true,linkcolor=blue,urlcolor=blue,filecolor=black,citecolor=red,pdfstartview=FitV,pdftitle={},pdfsubject={},pdfkeywords={},pdfpagemode=None,bookmarksopen=true]{hyperref}
\usepackage{graphicx}
\usepackage{amsmath}
\usepackage{amsfonts}
\usepackage{amssymb,ulem}
\usepackage{color}%
\usepackage{tikz}
\usepackage{dcolumn}
\usepackage{xcolor}

\newcommand{\red}[1]{{\color{red}{#1}}}

\begin{document}
\title{
  Entanglement Wedge Minimum Cross-section in Holographic Massive Gravity Theory
  }
\author{Peng Liu $^{1}$}
\email{phylp@email.jnu.edu.cn}
\author{Chao Niu $^{1}$}
\email{niuchaophy@gmail.com}
\author{Zi-Jian Shi $^{1}$}
\email{szj@stu2016.jnu.edu.cn}
\author{Cheng-Yong Zhang $^{1}$}
\email{zhangcy@email.jnu.edu.cn}
\thanks{corresponding author}
\affiliation{
  $^1$ Department of Physics and Siyuan Laboratory, Jinan University, Guangzhou 510632, China
}

\begin{abstract}

  We study the entanglement wedge cross-section (EWCS) in holographic massive gravity theory, in which a first and second-order phase transition can occur. We find that the mixed state entanglement measures, the EWCS and mutual information (MI) can characterize the phase transitions. The EWCS and MI show exactly the opposite behavior in the critical region, which suggests that the EWCS captures distinct degrees of freedom from that of the MI. More importantly, EWCS, MI and HEE  all show the same scaling behavior in the critical region. We give an analytical understanding of this phenomenon. By comparing the quantum information behavior in the thermodynamic phase transition of holographic superconductors, we analyze the relationship and difference between them, and provide two mechanisms of quantum information scaling behavior in the thermodynamic phase transition.

\end{abstract}
\maketitle
\tableofcontents

\section{Introduction}
\label{sec:introduction}

As a special property of the quantum system, quantum information plays an important role in other fields, such as condensed matter theory and holographic gravity theory. Many quantum phase transitions involving strong correlations could be characterized by entanglement measures \cite{Osterloh:2002na,Amico:2007ag,Wen:2006topo,Kitaev:2006topo}. Recently, holographic gravity theory has built a bridge between quantum information and geometry and has sparkled important insights into the geometric correspondence of quantum information properties and the understanding of the emergence of spacetime \cite{Ryu:2006bv,Hubeny:2007xt,Lewkowycz:2013nqa,Dong:2016hjy}.

There are many different measures of quantum entanglement, such as entanglement entropy (EE) and mutual information (MI), which characterize different properties of quantum systems. It is worth noting that EE, which is the most concerned physical quantity, is not suitable for describing the entanglement of mixed states, while the concepts of MI, R\'enyi entropy, entanglement of purification, reflected entropy and entanglement negativity are more suitable to describe the entanglement of mixed states \cite{vidal:2002,Horodecki:2009review}. A longstanding problem with quantum entanglement is that it is usually extremely difficult to calculate.

In recent years, holographic duality theory has been widely used to study strongly correlated physics, and it bridges the gap between geometry and quantum information. The earliest development is that the entanglement entropy of quantum field theory is proportional to the area of the minimum surface in dual gravity theory, which has been dubbed as the holographic entanglement entropy (HEE) \cite{Ryu:2006bv}. HEE has been proved a good diagnose of quantum phase transitions and thermodynamic phase transitions \cite{Nishioka:2006gr,Klebanov:2007ws,Pakman:2008ui,Zhang:2016rcm,Zeng:2016fsb,Ling:2015dma,Ling:2016wyr,Ling:2016dck,Kuang:2014kha,Guo:2019vni,Mahapatra:2019uql,Dudal:2018ztm,Dey:2015ytd}.
Subsequently, many other holographic duals of quantum information related physical quantities have been proposed, and their applications in strong correlation theory have been studied \cite{Nishioka:2006gr,Klebanov:2007ws,Pakman:2008ui,Zhang:2016rcm,Zeng:2016fsb}. For example, R\'enyi entropy has been proposed to be proportional to the area of the minimum cosmic brane \cite{Dong:2016fnf}. Entanglement of purification, reflected entropy, odd entropy and entanglement negativity have been proposed to be proportional to the area of the entanglement wedge minimum cross-section (EWCS) \cite{Takayanagi:2017knl,Nguyen:2017yqw,Kudler-Flam:2018qjo,Kusuki:2019zsp,Dutta:2019gen,Tamaoka:2018ned}. The EWCS provides a novel and powerful tool for studying the mixed state entanglement \cite{Yang:2018gfq,Ghodrati:2019hnn,Huang:2019zph,Fu:2020oep,Li:2021rff,Gong:2020pse,Liu:2019npm,Liu:2019qje,Lala:2020lcp,Bao:2018gck,Umemoto:2018jpc,Umemoto:2019jlz,Akers:2019gcv,Jain:2020rbb,Agon:2018lwq,Espindola:2018ozt,Basak:2020oaf,KumarBasak:2021lwm,BabaeiVelni:2019pkw,Saha:2021kwq,BabaeiVelni:2020wfl}. In addition, quantum complexity has been associated with the volume or action of certain region. Moreover, the butterfly velocity, a dynamical quantum information property of the quantum system, has been related to the geometry of the horizon of the black hole \cite{Shenker:2013pqa,Sekino:2008he,Maldacena:2015waa,Donos:2012js,Blake:2016wvh,Blake:2016sud,Ling:2016ibq,Ling:2016wuy,Wu:2017mdl,Liu:2019npm}. All these developments pave the way for the study of quantum information properties of strongly correlated systems in the framework of holographic duality theory.

Massive gravity plays an important role in the holographic duality theory because it can break the translational symmetry, thus producing momentum dissipation in the dual condensed matter system \cite{Vegh:2013sk,Blake:2013bqa,Blake:2013owa,Davison:2013jba,deRham:2014zqa,Baggioli:2014roa,Alberte:2015isw}. As an important theory of gravity, its entanglement property has been systematically studied \cite{Zeng:2015tfj}. However, the mixed state entanglement {property} in massive gravity theory has not been well studied. Therefore, the main goal of this paper is to study the mixed state entanglement in massive gravity --- the EWCS. In particular, we will discuss the properties of EWCS and its comparison with HEE and MI. It is noted that there is a thermodynamic phase transition {(Hawking-Page transition)} in massive gravity theory \cite{Vegh:2013sk,Cai:2014znn,Xu:2015rfa,Zeng:2015tfj}. Remind also that in the holographic superconductivity phase transition model, the EWCS has been found to have obvious non-smoothness at the critical points, thus diagnosing the thermal phase transitions \cite{Liu:2020blk}. Therefore, it is desirable to examine what role the mixed entanglement measures play during the phase transitions in massive gravity theory.
{More importantly, whether the quantum entanglement during superconductivity phase transition is different from that of Hawking-Page phase transition, and the underlying reasons are worth studying.}

We organize this paper as follows: we introduce the AdS massive gravity model in Sec. \ref{sec:alg}. We discuss the properties of HEE (\ref{sec:HEE}), MI (\ref{sec:mi}) and EWCS (\ref{sec:eop_phenomena}) systematically. In \ref{sec:critical}, we explore the scaling behavior of EWCS, MI and HEE. Finally, we summarize in Sec. \ref{sec:discuss}.

\section{Holographic Massive Gravity Theory}
\label{sec:alg}
The action of $n+2$-dimensional massive gravity system reads \cite{Cai:2014znn},
\begin{equation}\label{eq:actionn2}
  S = \frac { 1 } { 16\pi G_n } \int d ^ { n + 2 } x \sqrt { - g } \left[ R + \frac { n ( n + 1 ) } { L ^ { 2 } } - \frac { 1 } { 4 } F ^ { 2 } + m ^ { 2 } \sum _ { i } ^ { 4 } c _ { i } \mathcal { U } _ { i } ( g , f ) \right],
\end{equation}
where $G_n$ is the n-dimensional Newton constant which we set as $1$, $f_{\mu\nu}$ is the reference metric, $c_i$ are constants and $\mathcal{U}_i$ are symmetric polynomials of the eigenvalue of the $(n+2)\times(n+2)$ matrix $\mathcal { K } ^ { \mu } {_ { \nu }} \equiv \sqrt { g ^ { \mu \alpha } f _ { \alpha \nu } }$,
\begin{equation}\label{eq:matrixpoly}
  \begin{aligned} \mathcal { U } _ { 1 } & = [ \mathcal { K } ],\\
  \mathcal { U } _ { 2 } & = [ \mathcal { K } ] ^ { 2 } - \left[ \mathcal { K } ^ { 2 } \right], \\
  \mathcal { U } _ { 3 } & = [ \mathcal { K } ] ^ { 3 } - 3 [ \mathcal { K } ] \left[ \mathcal { K } ^ { 2 } \right] + 2 \left[ \mathcal { K } ^ { 3 } \right], \\
  \mathcal { U } _ { 4 } & = [ \mathcal { K } ] ^ { 4 } - 6 \left[ \mathcal { K } ^ { 2 } \right] [ \mathcal { K } ] ^ { 2 } + 8 \left[ \mathcal { K } ^ { 3 } \right] [ \mathcal { K } ] + 3 \left[ \mathcal { K } ^ { 2 } \right] ^ { 2 } - 6 \left[ \mathcal { K } ^ { 4 } \right]. \end{aligned}
\end{equation}
Here  $[\cdot]$ denotes the trace. The $F^2 \equiv F^{\mu\nu}F_{\mu\nu}$ is the square of the Maxwell field strength. $L$ is the AdS length scale, which we fix as $1$ for convenience. $m$ introduces mass to the graviton, which breaks the translational symmetry. The translational symmetry will be recovered when $m\to 0$.

The action \eqref{eq:actionn2} admits a solution \cite{Cai:2014znn},
\begin{equation}
  \label{eq:mgmetric}
  d s ^ { 2 } = - f ( r ) d t ^ { 2 } + f ^ { - 1 } ( r ) d r ^ { 2 } + r ^ { 2 } h _ { i j } d x ^ { i } d x ^ { j } , \quad i , j = 1,2,3 , \cdots , n.
\end{equation}
with the reference metric
\begin{equation}\label{eq:refmetric}
  f _ { \mu \nu } = \operatorname { diag } \left( 0,0 , c _ { 0 } ^ { 2 } h _ { i j } \right).
\end{equation}
The Maxwell field reads,
\begin{equation}\label{eq:maxwell}
  A_t = \mu - \frac { Q } { ( n - 1 ) r ^ { n - 1 } },
\end{equation}
where the chemical potential $\mu$ can be obtained by requiring a vanishing static electric potential,
\begin{equation}
  \label{eq:chemicalpo}
  \mu=\frac{Q}{(n-1) r_{+}^{n-1}}.
\end{equation}
The $\mathcal U_\#$ are
\begin{equation}\label{eq:allcs}
  \begin{aligned}
    \mathcal { U } _ { 1 } &= n c _ { 0 } / r , \\
    \mathcal { U } _ { 2 } &= n ( n - 1 ) c _ { 0 } ^ { 2 } / r ^ { 2 }, \\
    \mathcal { U } _ { 3 } &= n ( n - 1 ) ( n - 2 ) c _ { 0 } ^ { 3 } / r ^ { 3 } , \\
    \mathcal { U } _ { 4 } &= n ( n - 1 ) ( n - 2 ) ( n - 3 ) c _ { 0 } ^ { 4 } / r ^ { 4 },
  \end{aligned}
\end{equation}
and the function $f$ is
\begin{equation}
  \label{eq:fform}
  \begin{aligned}
    f(r)=& k+\frac{r^{2}}{L^{2}}-\frac{M}{r^{n-1}}+\frac{Q^{2}}{2 n(n-1) r^{2(n-1)}}+\frac{c_{0} c_{1} m^{2}}{n} r+c_{0}^{2} c_{2} m^{2} \\
    &+\frac{(n-1) c_{0}^{3} c_{3} m^{2}}{r}+\frac{(n-1)(n-2) c_{0}^{4} c_{4} m^{2}}{r^{2}}.
  \end{aligned}
\end{equation}
$M,\,Q$ are the mass and the charge of the black hole. $h_{ij}dx^idx^j$ is the line element of the Einstein space whose curvature is $n(n-1)k$. Therefore, we can denote the case of spherical, Ricci flat and hyperbolic horizon as $k=1,\,0,\,-1$. We consider the 4-dimensional flat case ($k=0$) in this paper, where $\mathcal U_3 = \mathcal U_4 =0$. Consequently, the last two terms in \eqref{eq:fform} vanishes. Note also that $m$ and $c_i$ in \eqref{eq:fform} are redundant, we denote $\alpha\equiv c _ { 0 } c _ { 1 } m ^ { 2 } / 2 ,\, \beta \equiv c _ { 0 } ^ { 2 } c _ { 2 } m ^ { 2 }$ such that
\begin{equation}
  \label{eq:fexps}
  f ( r ) =  - \frac { 2 M } { r } + \frac { Q ^ { 2 } } { 4 r ^ { 2 } } + r^2 + \alpha r + \beta.
\end{equation}

At the horizon $r=r_h$ we have $f(r_h)=0$. We can solve $M$ with $r_h$ and the Hawking temperature reads
\begin{equation}
  \label{eq:hawkingtemp}
  T= \frac{f'(r_h)}{4\pi} = -\frac{Q^2-4 r_h^2 \left(\beta +2 \alpha  r_h+3 r_h^2\right)}{16 \pi  r_h^3}.
\end{equation}
The effective entropy density \footnote{For $k=0$ case, the entropy diverges due to the infinitely large plane, therefore a density is adopted to discuss this problem. The entropy density, defined by dividing the entropy by the area is supposed to be $r^2_h/4$. However, in order to fasiliate the discussion of the thermadynamics, it is more convenient to define the effective entropy density such that $F=M-Ts$ can effectively capture the phase transitions. See \cite{Cai:2014znn} for a more complete formalism.} is $s \equiv \pi r^2_h$. We can rewrite the temperature as
\begin{equation}
  \label{eq:tement}
  T=\frac{-\pi ^2 Q^2+8 \sqrt{\pi } \alpha  s^{3/2}+12 s^2+4 \pi  \beta  s}{16 \pi ^{3/2} s^{3/2}}.
\end{equation}
Solving $\partial_s T =0$ we have two roots,
\begin{equation}
  \label{eq:roots}
  s = \frac{\pi}{6} \left(\beta \pm \sqrt{\beta ^2-9 Q^2}\right).
\end{equation}
From the above equation, we {can} find that the van der Waals phase transition occurs when the above two roots are all positive, i.e.,
\begin{equation}
  \label{eq:vander}
  \beta \geqslant 3Q.
\end{equation}
We are particularly interested in the phase transition process. The first-order phase transition $(\beta > 3Q)$ occurs {when} the entropy density jumps. Meanwhile, the second-order phase transition $(\beta = 3Q)$ occurs  {when} the entropy density is continuous while its first derivative to temperature is discontinuous.

\begin{figure}
  \centering
  \includegraphics[width=0.7\textwidth]{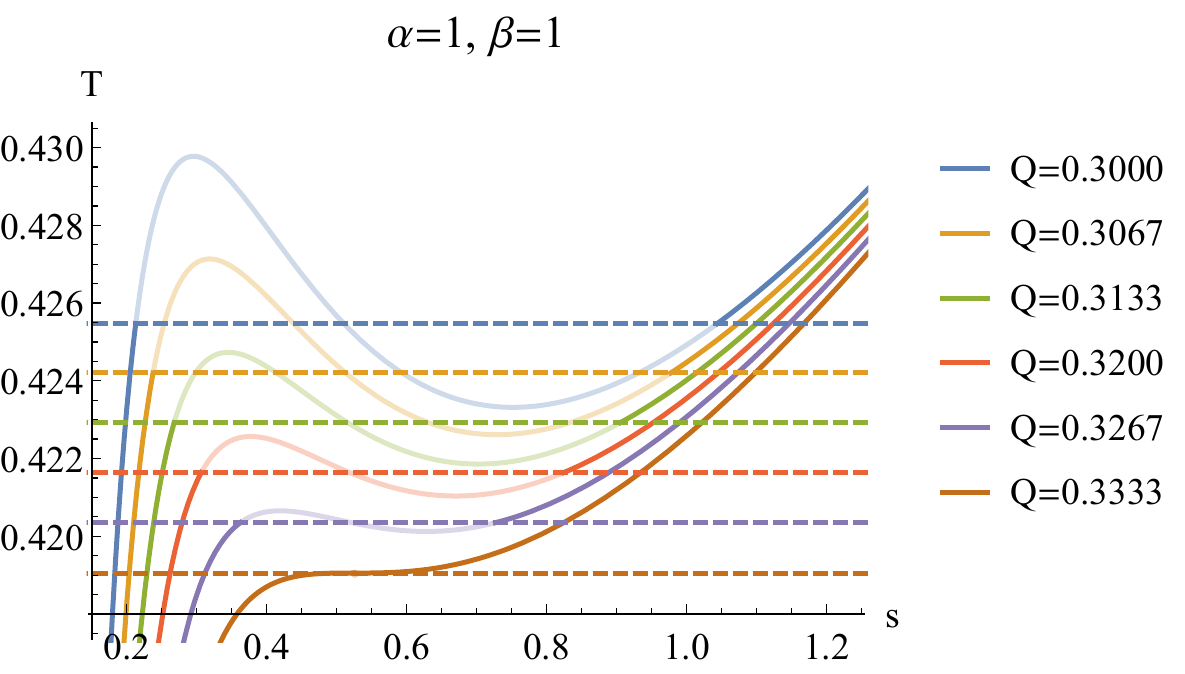}
  \caption{The phase transition at several different values of the $Q$. The solid lines are the $T-s$ relation at $Q$ specified by the plot legends, where the lighter segments correspond to the metastable regions. The dashed horizontal lines, of which the parameters match the color of the plot legends, are the critical temperature where the phase transition occurs. The brown curve $ Q=0.3333 $ corresponds to the critical case of the second-order phase transition.}
  \label{fig:phasetransition}
\end{figure}
We show an example of $(\alpha,\beta)=(1,1)$ in Fig. \ref{fig:phasetransition}, from which we can find that the system undergoes a van der Waals like thermal phase transition for $Q<1/3$. At $Q=1/3$, the system undergoes a second-order phase transition. Further increasing the $Q${, the thermal phase transition is absent}. Near the critical point $\beta = 3Q$ where  second-order phase transition occurs, as can be seen from Fig. \ref{fig:phasetransition}, there is
\begin{equation}\label{eq:dsdt}
  s'(T) \to \infty.
\end{equation}
This suggests that a critical scaling behavior will emerge in the critical region. The entropy density and temperature at the critical point are respectively
\begin{equation}\label{eq:critcalst}
  s_c = \frac{\pi Q}{2},\quad T_c = \frac{\alpha +2 \sqrt{2Q}}{2 \pi }.
\end{equation}
Near the critical point, there is \begin{equation}\label{eq:svst}
  (s-s_c) \sim (T-T_c)^{\alpha_s},
\end{equation}
in which the exponent $\alpha_s$ is called the critical exponent of the entropy density. Given the temperature expression, one can obtain
\begin{equation}\label{eq:dsdtexp}
  s'(T) = \frac{32 \pi ^{3/2} s^{5/2}}{3 \pi ^2 Q^2+12 s^2-4 \pi  \beta  s}.
\end{equation}
Expanding \eqref{eq:dsdtexp} near $s=s_c$ as
\begin{equation}\label{eq:dsdtexpexpand}
  s'(T) =\frac{\sqrt{2} \pi ^4 Q^{5/2}}{3 \left(s-\frac{\pi  Q}{2}\right)^2}+\frac{5 \sqrt{2} \pi ^3 Q^{3/2}}{3 \left(s-\frac{\pi  Q}{2}\right)}+\frac{5 \pi ^2 \sqrt{Q}}{\sqrt{2}}+O\left(\left(s-\frac{\pi  Q}{2}\right)^1\right),
\end{equation}
 we find that the leading order is $s'(T) \sim (s-s_c)^{-2}$. Together with \eqref{eq:svst}, one can immediately obtain the critical exponent as
\begin{equation}\label{eq:exponents}
  \alpha_s = 1/3.
\end{equation}
Similar analysis can also be found in \cite{Zeng:2015tfj}.

\section{The Holographic entanglement entropy}\label{sec:HEE}
Now we study the entanglement structure of the system. Entanglement, as an important feature that distinguishes quantum systems from classical systems, can be described by many physical quantities. One of the most famous is the EE. For a system composed of $A$ and $B$, the properties of $A$ are described by a reduced density matrix $\rho_A = \text{Tr}_{B} \rho_{\text{total}}$ where $\rho_{\text{total}}$ is the density matrix of the whole system. EE is defined as the von Newmann entropy of the reduced density matrix $\rho_A$,
\begin{equation}\label{ee-von}
  S_{A} (|\psi\rangle) = - \text{Tr}\left[ \rho_{A} \log \rho_{A} \right].
\end{equation}
For a system in pure state $|\psi\rangle$, $\rho_{A} = \text{Tr}_{B} \left(|\psi\rangle\langle\psi|\right)$. The above definition leads to $S_A = S_B$ \cite{Chuang:2002book}. EE has been widely recognized as a good entanglement measure for pure states. However, it is not suitable for describing the entanglement of mixed states. The reason is that the degrees of freedom in $A$ and $B$ in the direct product state $\rho_A \otimes \rho_B$ are not entangled, but they can have non-zero EE. Several new entanglement measures have been proposed to characterize the entanglement of mixed states, among which MI is the most commonly used one \cite{vidal:2002,Horodecki:2009review}.
\begin{figure}
  \begin{tikzpicture}[scale=1]
    \node [above right] at (0,0) {\includegraphics[width=7.5cm]{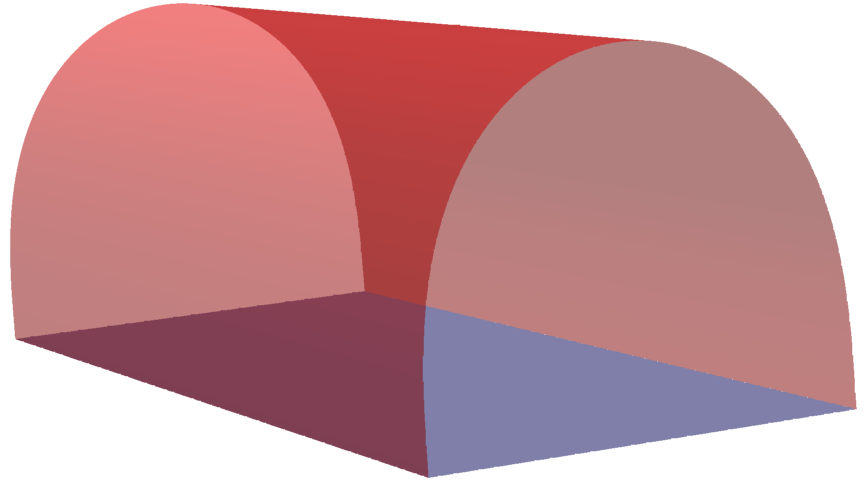}};
    \draw [right,->,thick] (3.85, 0.22) -- (6.25, 0.58) node[below] {$x$};
    \draw [right,->,thick] (3.85, 0.22) -- (1.25, 1.08) node[below] {$y$};
    \draw [right,->,thick] (3.85, 0.22) -- (3.7, 3.125) node[above] {$z$};
  \end{tikzpicture}
  \begin{tikzpicture}[scale=1]
    \node [above right] at (0,0) {\includegraphics[width=7.5cm]{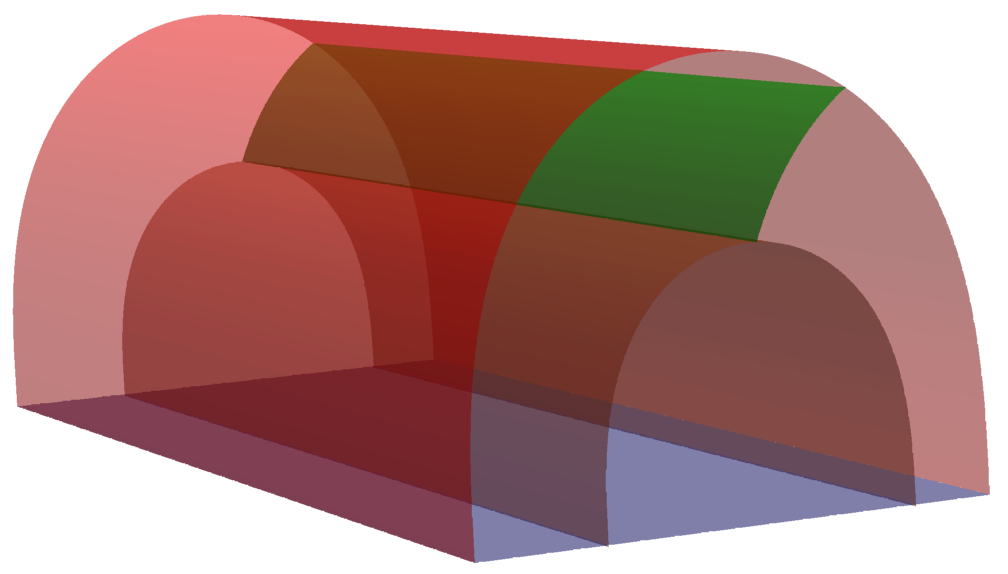}};
    \draw [right,->,thick] (3.67, 0.22) -- (6.25, 0.55) node[below] {$x$};
    \draw [right,->,thick] (3.67, 0.22) -- (1.25, 1.05) node[below] {$y$};
    \draw [right,->,thick] (3.67, 0.22) -- (3.6, 3.125) node[above] {$z$};
  \end{tikzpicture}

  \caption{The left plot: The minimum surface for a given width $w$. The right plot: The minimum cross-section (green surface) of the entanglement wedge.}
  \label{msd1}
\end{figure}
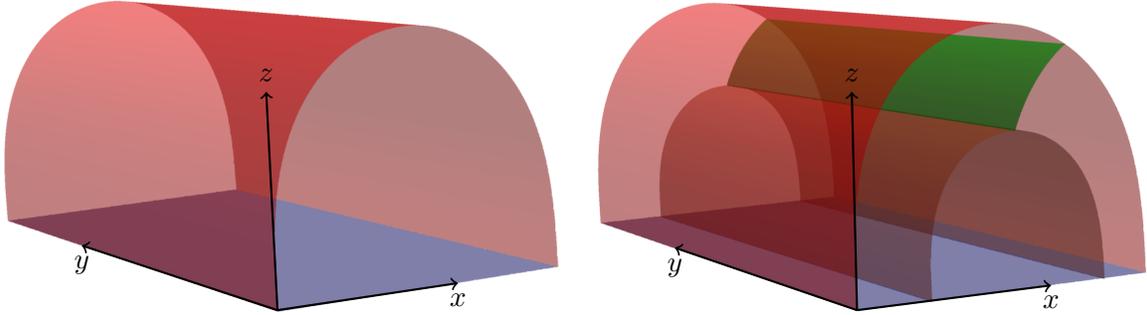

In holographic duality theory, the EE is related to the minimum area stretching into the bulk of the dual gravity systems \cite{Ryu:2006bv} (see Fig. \ref{msd1}). Here, we consider the infinite strip along $y$-direction, thus the minimum surfaces will be invariant along $y$-direction. {Adopting the angle as the parameterization,}
the minimum surface can be solved efficiently (see Fig. \ref{fig:cartoon4eop}). In this method, the range of angle is $(0, \pi/2)$. The first step of our numerical method is to discretize the angle with Gauss-Lobatoo collocation \cite{Boyd:2001}. Because the equation\red{s} of motion for the minimal surface are nonlinear, we  need to apply Newton-Raphson iteration method to find the  {minimum} surface based on  {the} discretization.

\begin{figure}[]
  \centering
  \includegraphics[width =0.6\textwidth]{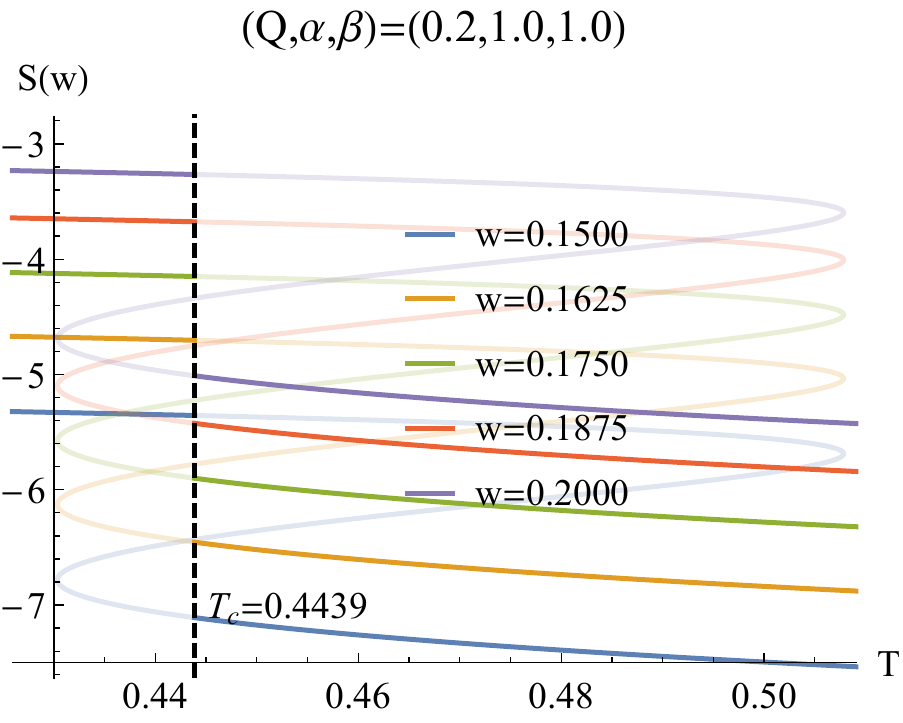}
  \caption{HEE vs $T$ at several small widths, where the lighter segments correspond to the metastable regions. The red dashed horizontal line represents the critical temperature $T_c = 0.4439$ of the first-order phase transition.}
  \label{fig:HEEvst1}
\end{figure}

First, we discuss the HEE during the first-order phase transition {where} $ (Q,\alpha,\beta)=(0.2,1.0,1.0) $. We find that the relationship between the   HEE and temperature is  {depends on the configuration}, which is completely different from other black holes, such as AdS-RN black holes \cite{Ling:2015dma}. Fig. \ref{fig:HEEvst1} is the HEE at small configurations, from which we can see that the HEE decreases with increasing temperature. The segments with lighter colors correspond to  the metastable regions. Moreover, by comparing the HEE at different widths (different curves) in this figure we can find that the HEE increases with the increase of the width $w$. The black dotted line in the figure shows the critical temperature $T_c = 0.4439$ of the first-order phase transition. When the temperature drops to the critical temperature, HEE will jump abruptly.
When crossing the critical point of the first-order phase transition, as shown in Figure 2, the horizon radius $r_h$ of the black hole will jump abruptly. Therefore, the sudden jump of geometry produces the sharp jump of the HEE.

\begin{figure}[]
  \centering
  \includegraphics[width =0.6\textwidth]{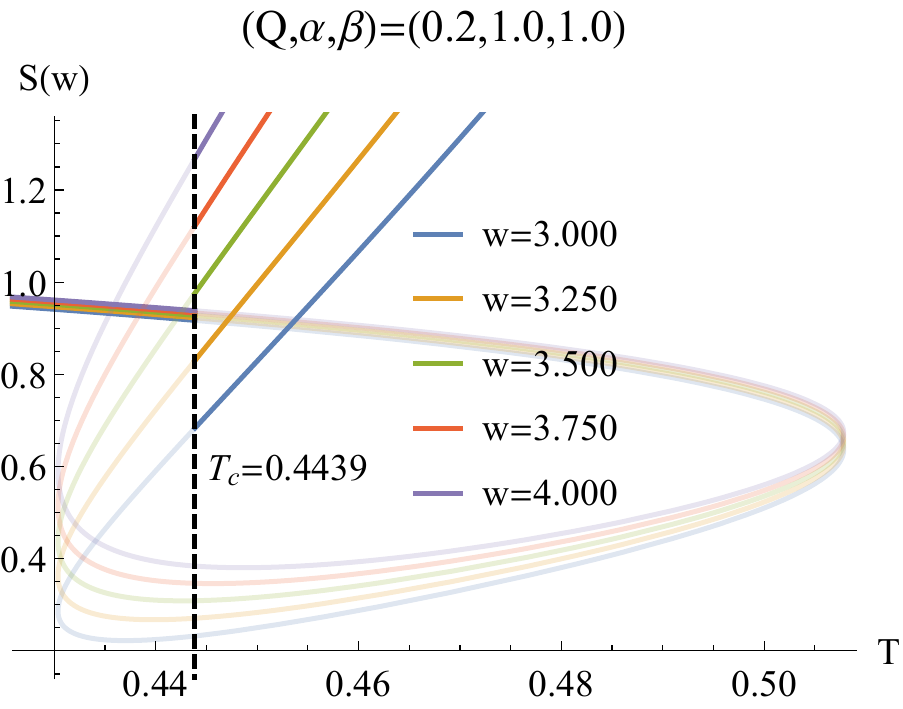}
  \caption{
    HEE vs $T$ at several large widths specified by the plot legends, where the lighter segments correspond to the metastable regions. The black dashed vertical line represents the critical temperature $T_c = 0.4439$ of the first-order phase transition.
    For each solid curve, the segments with lighter color are the unstable regions.
  }
  \label{fig:HEEvst2}
\end{figure}
When the configuration size increases, HEE gradually presents a more complicated phenomenon. It can be seen from Fig. \ref{fig:HEEvst2} that with the decrease of temperature, HEE decreases first. But when the temperature  {drops below} the critical temperature, the HEE increases with the decrease of temperature.
{Next, we give an understanding of these phenomena.}
Firstly, the decrease of HEE with the decrease of temperature in a large configuration can be understood as the contribution of thermodynamic entropy. In Fig. \ref{fig:phasetransition}, the entropy density decreases rapidly with the decrease of temperature above the critical temperature. Therefore, the entanglement entropy decreases rapidly above the critical temperature. However, when the temperature drops below the critical temperature, it can be seen from Fig. \ref{fig:phasetransition} that the entropy density changes very slowly with temperature. The behavior of entropy density with temperature no longer dominates the change of HEE with temperature, instead, the whole bulk geometry controls the change of HEE with temperature.

\begin{figure}
  \centering
  \includegraphics[width=0.45\textwidth]{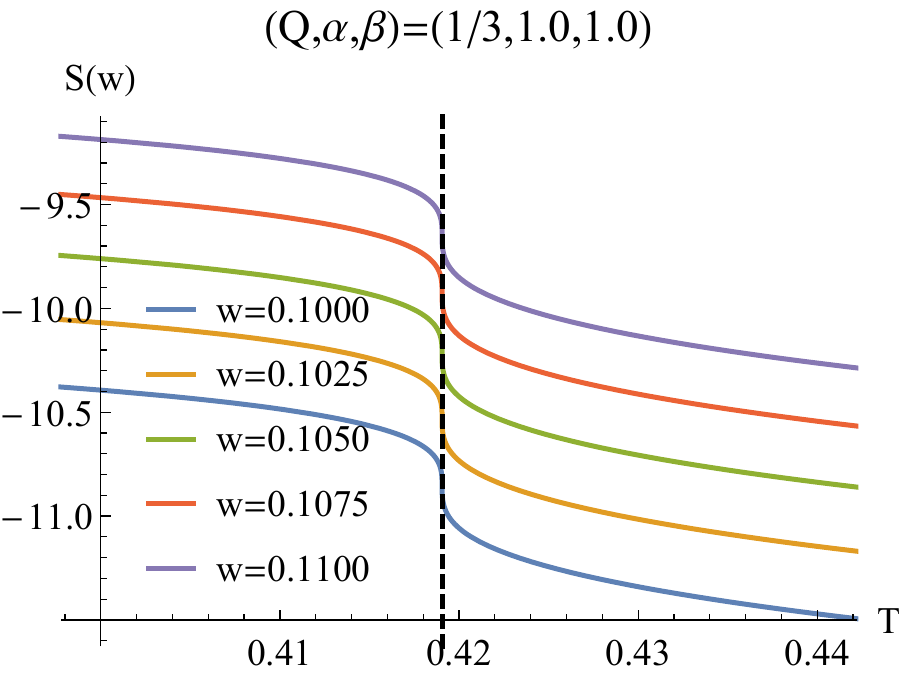}\quad
  \includegraphics[width=0.45\textwidth]{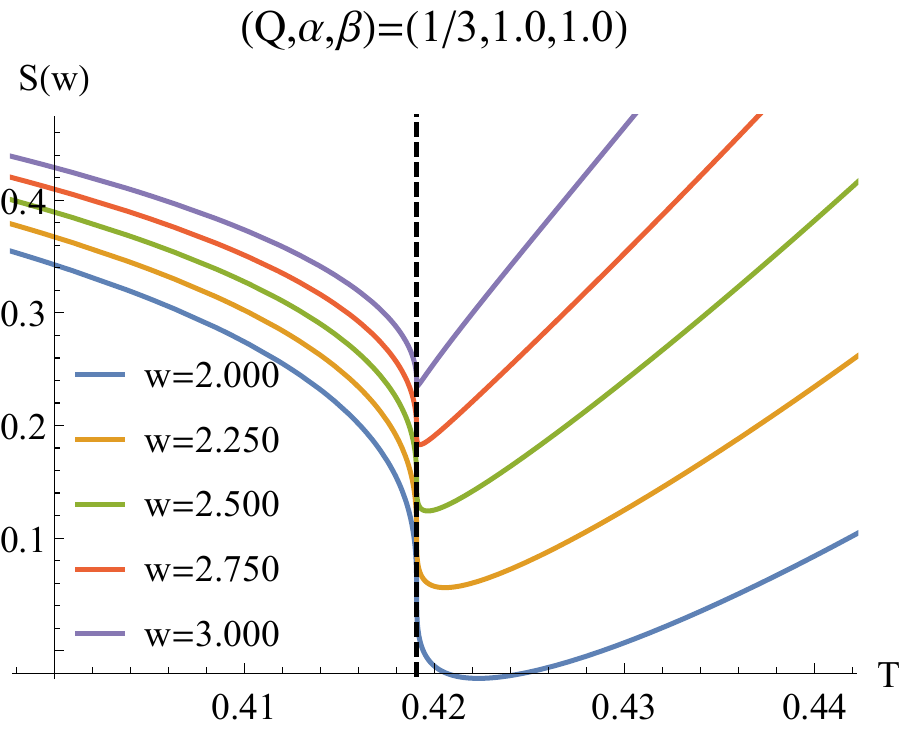}
  \caption{ The relationship between the temperature and the HEE for small (left plot) and large widths (right plot) at the critical case with second-order phase transition. The red dashed horizontal line represents the critical temperature $T_c = 0.4191$ of the second-order phase transition.}
  \label{fig:2-ndphasetransition}
\end{figure}

Now, we {study} the second-order phase transition with $(Q,\alpha,\beta)=(1/3,1,1)$. Regardless of the configuration, HEE exhibits a singular behavior at the critical temperature (see Fig. \ref{fig:2-ndphasetransition}). This is the response of HEE for the second-order phase transition. The HEE behavior at the second-order phase transition point is still related to the specific configuration. When the configuration is small (left figure in Fig. \ref{fig:2-ndphasetransition}), HEE increases with the decrease of temperature. However, when the configuration is large (right figure in Fig. \ref{fig:2-ndphasetransition}), HEE first decreases with the decrease of temperature, but increases after passing the critical point. The dependence of HEE on configuration size before and after the second-order phase transition can also be understood by comparing the contribution of thermodynamic entropy with that of whole bulk geometry (see the similar discussion in the previous paragraph).

\begin{figure}
  \centering
  \includegraphics[width=0.45\textwidth]{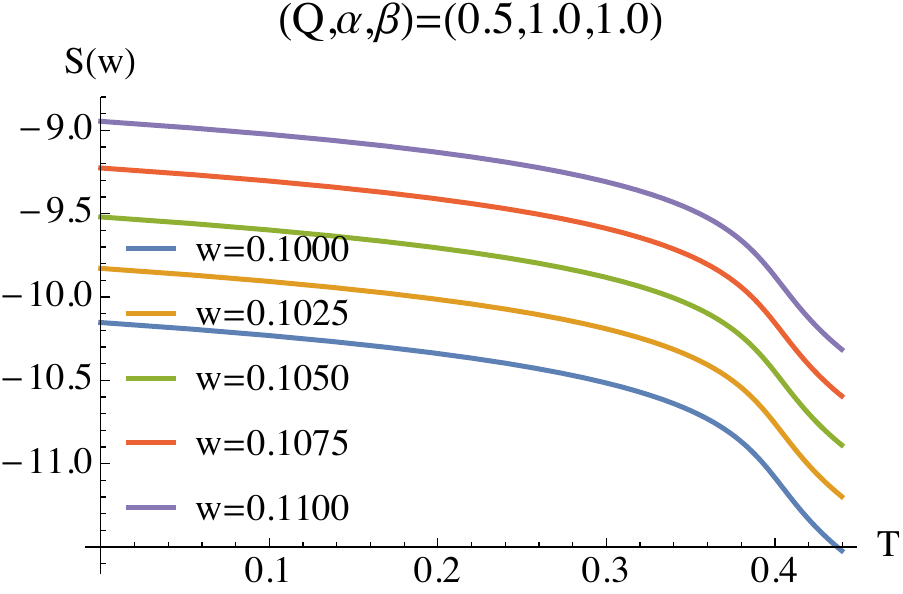}\quad
  \includegraphics[width=0.45\textwidth]{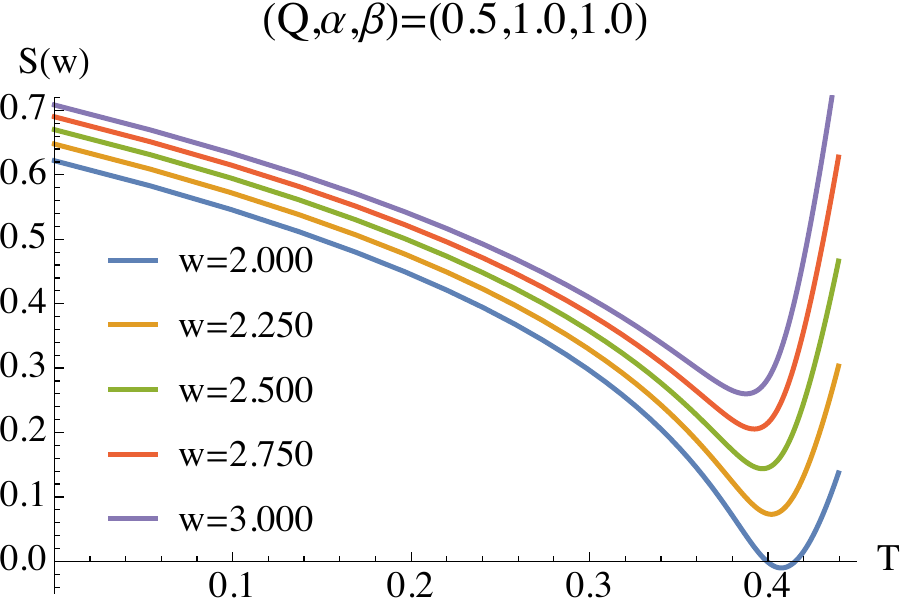}
  \caption{ The relationship between the temperature and the HEE for small (left plot) and large widths (right plot) in absence of thermal phase transitions. }
  \label{fig:nophased}
\end{figure}

Finally, we discuss the case without phase transition with $(Q,\alpha,\beta)=(0.5,1,1)$. From Fig. \ref{fig:nophased} we find that, when the configuration is small, HEE increases with the decrease of temperature. However, when the configuration is large, the HEE decreases first and then increases with the decrease of temperature. These phenomena are quite different from other black holes, such as AdS-RN black holes.

The above phenomena indicate that the HEE behavior in the massive gravity model can diagnose the first and second-order phase transitions, and  {the behavior of the} HEE is closely related to  specific configurations. Only when the temperature is relatively high or the configuration is relatively large, the HEE will increase with the temperature. This is as expected, because a higher temperature or larger {subregion} will render the minimum surface closer to the event horizon of the black hole, and its HEE will be dominated by {thermodynamic} entropy.

After elaborating on the properties of HEE, we  {then} discuss the properties of MI, one of the entanglement measures of mixed states.

\section{The holographic mutual information}\label{sec:mi}

\begin{figure}
  \centering
  \includegraphics[width=0.5\textwidth]{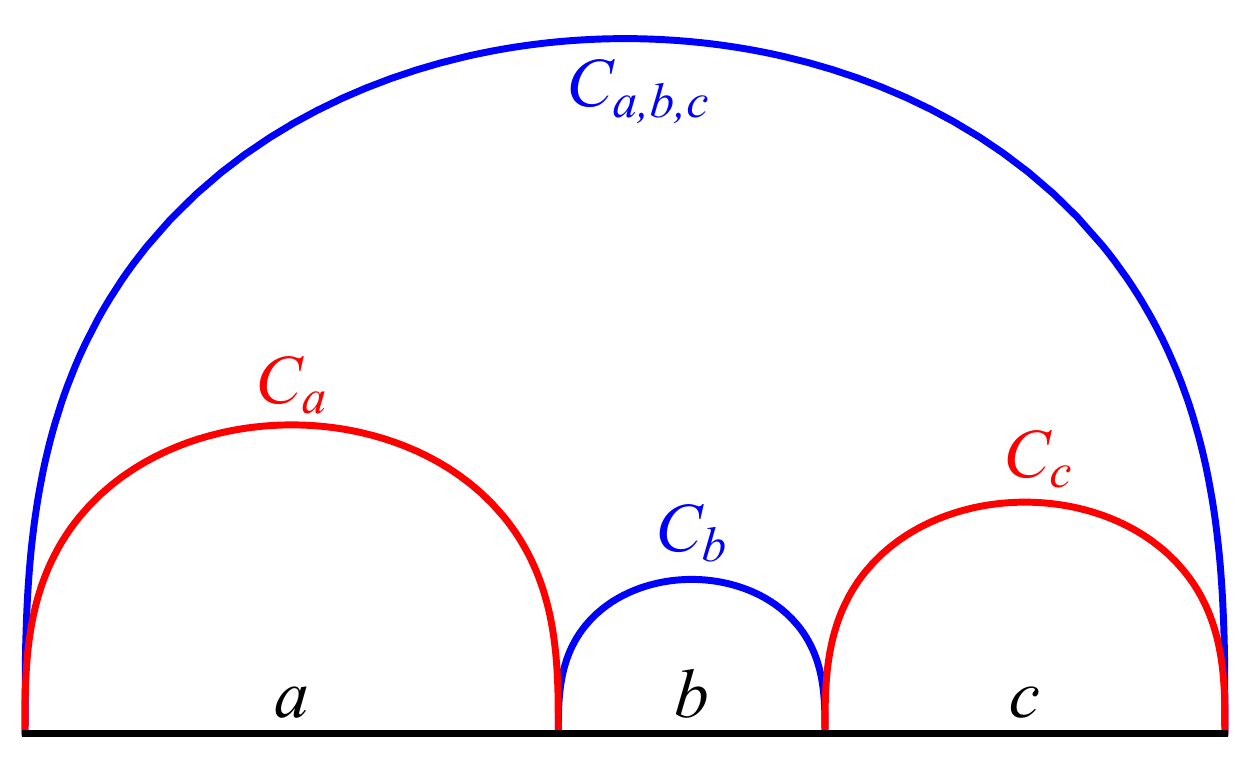}
  \caption{The demonstration of mutual information.}
  \label{fig:midemo}
\end{figure}

For disjoint subregion $A\cup C$ with separation $B$, the MI is defined as
\begin{equation}\label{mi:def}
  I\left(A,C\right) := S\left(A\right) + S\left(C\right) - S\left(A\cup C\right),
\end{equation}
which can measure the entanglement between $A$ and $C$. It is then straightforward to verify that $ I\left(A,C\right) =0 $ when $\rho_{AC} = \rho_{A} \otimes \rho_{C} $.
Therefore, MI can recognize that the direct product state has no entanglement. Because the definition of MI is closely related to HEE, we can directly {study} the properties of MI   based on HEE calculation.  The {minimum surface corresponding to} $S(A\cup C)$ has two  {candidates}, the red {curves} $C_a\cup C_c$  and blue {curves} $C_b\cup C_{a,b,c}$  in Fig. \ref{fig:midemo}, in which we have labeled the width of $A,\,B,\,C$ with $a,b,c$, respectively.

\begin{figure}
  \centering
  \includegraphics[width=0.6\textwidth]{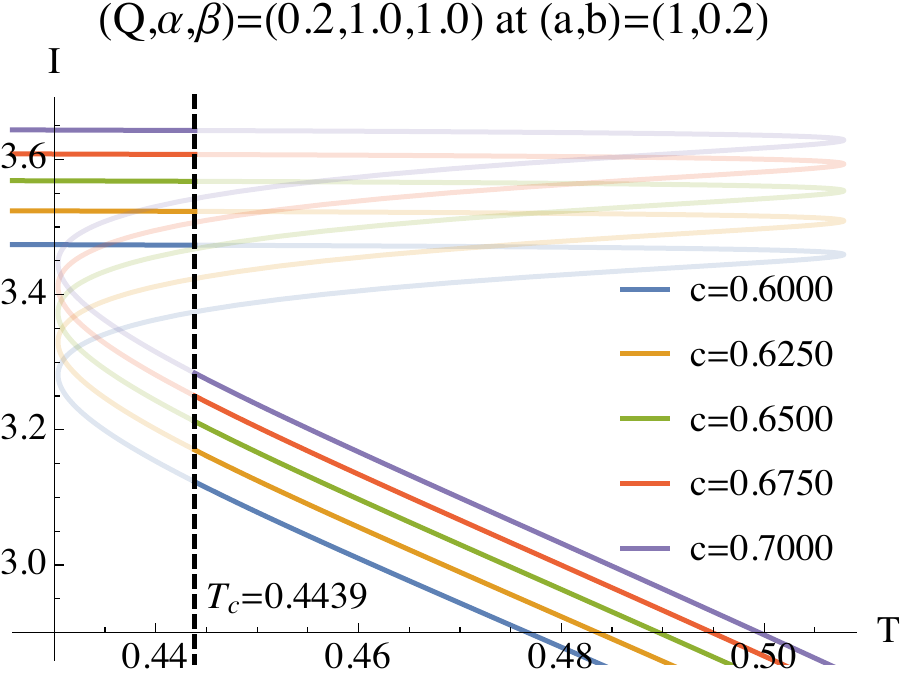}
  \caption{The $T$ vs $I$ in the scenario with a first-order phase transition, where the lighter segments correspond to the metastable regions. The red dashed horizontal line represents the critical temperature $T_c = 0.4439$ of the first-order phase transition.}
  \label{fig:firstordermi}
\end{figure}

First, we focus on the first-order phase transition at $(Q,\alpha,\beta)=(0.2,1.0,1.0)$. As can be seen from Fig. \ref{fig:firstordermi}, when the configuration is small, the behavior of MI with temperature is similar to that of HEE  (see Fig.\ref{fig:HEEvst1}). This behavior can be derived from the definition of the relationship between MI and HEE.
Since the HEE of small configuration changes more rapidly with temperature (as can be seen by comparing HEE at the small and large configuration in Fig. \ref{fig:nophased}), the behavior of MI with temperature is dominated by the part with the minimum width. Therefore, MI in small configuration is qualitatively consistent with that of HEE. It is noteworthy that, unlike the apparent configuration-dependent behavior of HEE, these phenomena of MI are configuration-independent.

\begin{figure}
  \centering
  \includegraphics[width=0.6\textwidth]{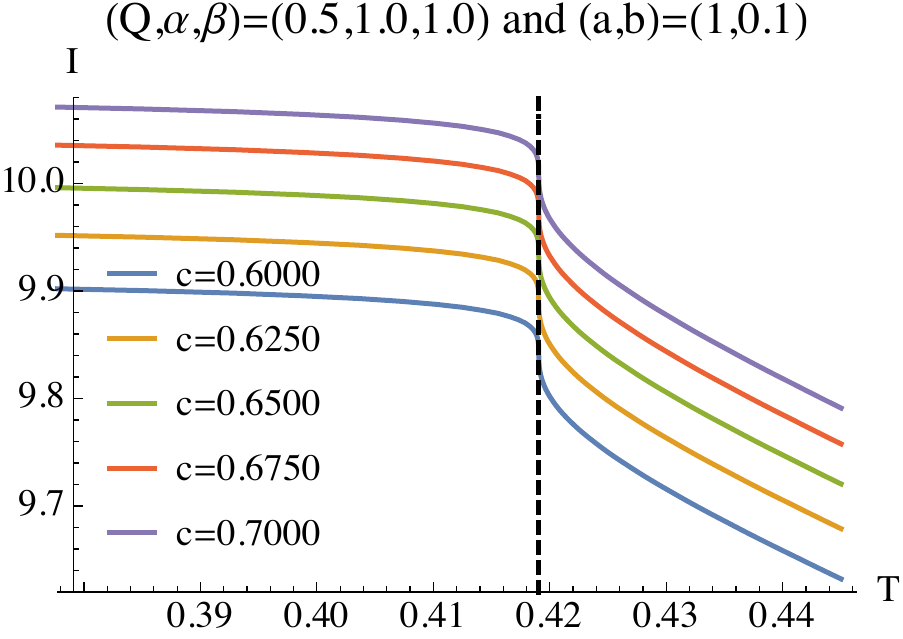}
  \caption{The $T$ vs $I$ in the scenario with a second-order phase transition. The red dashed horizontal line represents the critical temperature $T_c = 0.4191$ of the second-order phase transition.}
  \label{fig:criticalmi}
\end{figure}

Now, we study the behavior of MI when the second-order phase transition occurs. First, similar to the monotonic behavior at the first-order phase transition, MI again increases with decreasing temperature (see Fig. \ref{fig:criticalmi}). Moreover, the behavior of MI becomes singular at the critical temperature of the second-order phase transition. This is the reflection of the second-order phase transition. In addition, by comparing  {the MI at different $c$'s (different curves)}, we can see that MI always increases with the increase of $c$. This phenomenon is in line with expectations because larger configurations mean more entanglement between  {separate subregions}.

\begin{figure}
  \centering
  \includegraphics[width=0.6\textwidth]{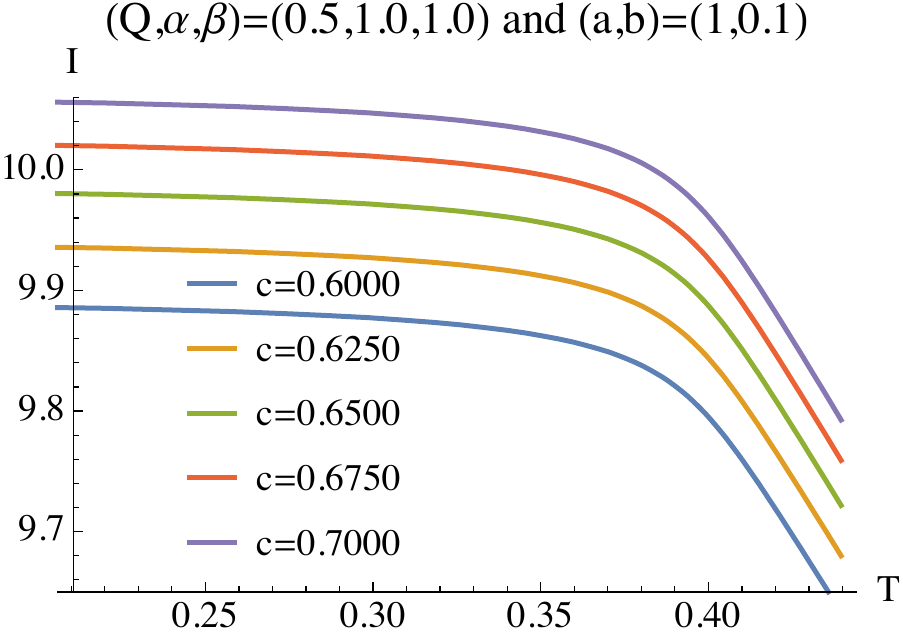}
  \caption{The $T$ vs $I$ in the case without thermal phase transitions.}
  \label{fig:noptmi1}
\end{figure}

Finally, we  {study} the behavior of MI in the absence of phase transition. As can be seen from Fig. \ref{fig:noptmi1}, that MI increases monotonically with decreasing temperature. This phenomenon is consistent with the cases in presence of phase transitions.

We summarize the behavior of MI in massive gravity. Firstly, MI can reflect the first-order and second-order phase transitions of massive gravity, with discontinuous and singular behavior at the critical temperatures, respectively. Secondly, the MI {increases} with decreasing temperature,  independent of the specific configuration or system parameter. From the above phenomena, we can see that MI is directly determined by HEE in certain cases. Therefore, MI is not a perfect measure of entanglement of mixed States.  {W}e need to resort to other mixed-state entanglement measures. Next, we study the EWCS, a novel mixed state entanglement measure.

\section{The entanglement wedge minimum cross-section}\label{sec:eop_phenomena}

Recently, several entanglement measures related to the purification process have attracted extensive attention, such as entanglement of purification, reflected entropy and so on \cite{Kudler-Flam:2018qjo,Kusuki:2019zsp,Dutta:2019gen}. The measure of entanglement of these mixed states has been proposed  dual to EWCS, which indicates that EWCS is a widely accepted geometric dual of entanglement of mixed states. Takayanagi proposed that the entanglement of purification $ E_{W}\left(\rho_{AB}\right) $ is proportional to the area of the EWCS $ \Sigma_{AB} $ \cite{Takayanagi:2017knl},
\begin{equation}\label{heop:def}
  E_{W}\left(\rho_{AB}\right) = \min_{\Sigma_{AB}} \left( \frac{\text{Area} \left(\Sigma_{AB}\right)}{4G_{N}}\right) .
\end{equation}
When the MI is zero, the entanglement wedge becomes disconnected, and hence EWCS vanishes because the cross-section does not exist.

Solving EWCS is difficult for three-fold reasons. First, the equation{s} of motion for solving the minimal surface {are} highly nonlinear. Second, the minimum cross-section, which minimizes the cross-section with each cross-section itself a local minimum surface, is usually hard to find. Last but not least, in the presence of the black hole, the numerical precision is easily sabotaged by the coordinate singularity near the black hole horizon.

\begin{figure}
  \centering
  \includegraphics[width = \textwidth]{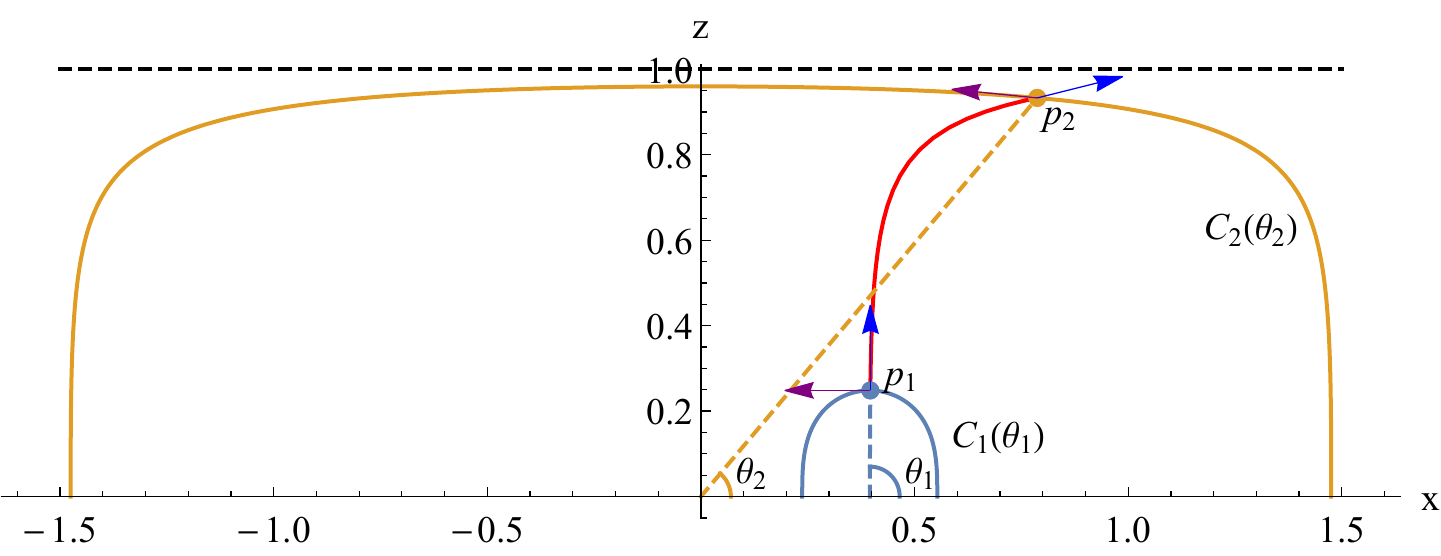}
  \caption{
  The schematic demonstration of the EWCS. The $p_1$ and $p_2$ are the intersection of the minimum surface connecting those two minimum surfaces. The solid blue curve (parametrized by $\theta_1$) and solid orange curve (parametrized by $\theta_2$) are minimum surfaces. The thick purple curve is the minimum surface connecting $p_1$ and $p_2$. The blue arrows at the $p_1$ and $p_2$ are the tangent vectors $\left.\left(\frac{\partial}{\partial z}\right)^a\right|_{p_1}$ and $\left.\left(\frac{\partial}{\partial z}\right)^a\right|_{p_2}$ along the $C_{p_1,p_2}$, while the purple arrows are the tangent vectors $\left.\left(\frac{\partial}{\partial \theta_1}\right)^a\right|_{p_1}$ and $\left.\left(\frac{\partial}{\partial \theta_2}\right)^a\right|_{p_2}$ along $C_1,\,C_2$, respectively. The dark dashed horizontal line represents the black brane horizon.}
  \label{fig:cartoon4eop}
\end{figure}

Recently, we proposed an efficient algorithm for solving EWCS by using the property that the minimum cross-section must be locally {perpendicular} to the  {boundaries of the} entanglement wedge. Fig. \ref{fig:cartoon4eop} is a schematic diagram of the main ideas of the algorithm for solving EWCS. Considering the EWCS of two parallel infinite strips along $y$-direction in a homogeneous background,
\begin{equation}\label{genbg}
  ds^{2} = {g_{tt}} dt^2 + g_{zz}dz^2 + g_{xx}dx^2 + g_{yy} dy^2,
\end{equation}
where $z=0$ represents the asymptotic AdS boundary. Due to the translational invariance along $y$-direction, the EWCS will also be invariant along $y$-direction. The homogeneity implies that metric components $g_{\mu\nu}$ are functions of $z$ only.
For a biparty subsystem with minimum surfaces $C_1(\theta_1),\,C_2(\theta_2)$, we work out the minimum surface $C_{p_1,p_2}$ connecting $p_1 \in C_1$ and $p_2\in C_2$. We can parametrize $C_{p_1,p_2}$ with $z$, and hence the area of $C_{p_1,p_2}$ reads,
\begin{equation}\label{eq:zpara}
  A = \int_{C_{p_1,p_2}} \sqrt{ g_{xx} g_{yy} x'(z)^2 + g_{zz} g_{yy} } dz.
\end{equation}
The equation of motion reads,
\begin{equation}\label{eq:zparaeom}
  x'(z)^3 \left(\frac{g_{ xx } g_{ yy }'}{2 g_{ yy } g_{ zz }}+\frac{g_{ xx }'}{2 g_{ zz }}\right)+x'(z) \left(\frac{g_{ xx }'}{g_{ xx }}+\frac{g_{ yy }'}{2 g_{ yy }}-\frac{g_{ zz }'}{2 g_{ zz }}\right)+x''(z) =0,
\end{equation}
with boundary conditions,
\begin{equation}\label{eq:zparabcs}
  x(z(\theta_i)) = x(\theta_i),\quad i=1,2.
\end{equation}
The local  {perpendicular condition} between the minimum cross-section and the entanglement wedge implies that
\begin{equation}\label{eq:perpend}
  \left\langle \frac{\partial}{\partial z},\frac{\partial}{\partial \theta_1} \right\rangle_{p_1} = 0,\quad \left\langle \frac{\partial}{\partial z},\frac{\partial}{\partial \theta_2} \right\rangle_{p_2} = 0,
\end{equation}
where $\langle\cdot,\cdot\rangle$ represents the vector product {measured by} the metric $g_{ab}$. For better numerical stability control, we adopt the normalized local orthogonal relation,
\begin{equation}\label{eq:perpend2}
    Q_1(\theta_1,\theta_2) \equiv \left.\frac{ \left\langle \frac{\partial}{\partial z},\frac{\partial}{\partial \theta_1} \right\rangle }{\sqrt{ \left\langle \frac{\partial}{\partial z},\frac{\partial}{\partial z} \right\rangle \left\langle \frac{\partial}{\partial \theta_1},\frac{\partial}{\partial \theta_1} \right\rangle }}\right|_{p_1} = 0, \quad
    Q_2(\theta_1,\theta_2)  \equiv \left.\frac{ \left\langle \frac{\partial}{\partial z},\frac{\partial}{\partial \theta_1} \right\rangle }{\sqrt{ \left\langle \frac{\partial}{\partial z},\frac{\partial}{\partial z} \right\rangle \left\langle \frac{\partial}{\partial \theta_2},\frac{\partial}{\partial \theta_2} \right\rangle }}\right|_{p_2} = 0.
\end{equation}
Finding the EWCS is turned to locating the minimum surface anchoring at $(\theta_1,\theta_2)$ where \eqref{eq:perpend2} is satisfied. To this end, we adopt the Newton-Raphson method to locate the endpoints satisfying the local perpendicular conditions.  Based on the above techniques, we now study the relationship between the thermal phase transition and the EWCS.

\begin{figure}
  \centering
  \includegraphics[width=0.6\textwidth]{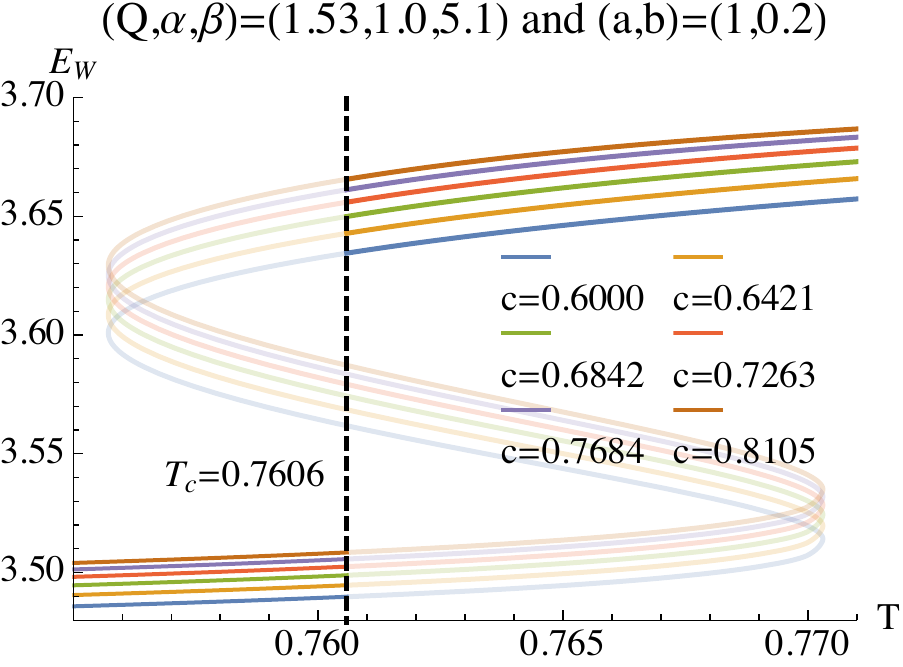}
  \caption{The $T$ vs EWCS in presence of the first-order phase transition, where the lighter segments correspond to the metastable regions. The red dashed horizontal line represents the critical temperature $T_c = 0.7547$ of the first-order phase transition.}
  \label{fig:eop1st}
\end{figure}

We show the relationship between the EWCS and temperature during the first-order phase transition in Fig. \ref{fig:eop1st}. It can be seen from the figure that the first-order phase transition does have an impact on EWCS. The EWCS also exhibits a van der Waals-like behavior. As the temperature decreases, the EWCS decreases and jumps at the critical temperature of the first-order phase transition. Note that MI increases with decreasing temperature (Fig. \ref{fig:firstordermi}), and the {the EWCS exhibits the opposite behavior}. These phenomena of EWCS are also independent of {configurations}.

\begin{figure}
  \centering
  \includegraphics[width=0.6\textwidth]{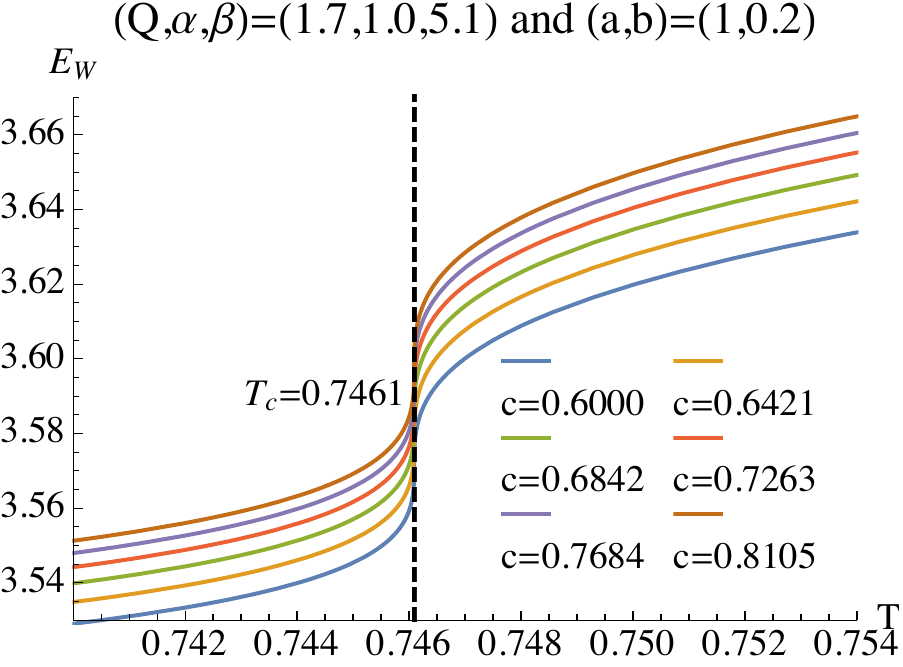}
  \caption{The temperature vs EWCS at the second-order phase transition. The phenomenon is qualitatively the same for other configurations and parameters. The red dashed horizontal line represents the critical temperature $T_c = 0.7461$ of the second-order phase transition.}
  \label{fig:eop2nd}
\end{figure}

Now, we focus on the  second-order phase transition. It can be seen from Fig. \ref{fig:eop2nd} that, similar to MI, the EWCS also exhibits a singular behavior at the phase transition. The difference is that here again the behavior of EWCS with temperature is opposite to that of MI. EWCS decreases with decreasing temperature, while MI increases with decreasing temperature (Fig.  \ref{fig:criticalmi}).

\begin{figure}
  \centering
  \includegraphics[width=0.6\textwidth]{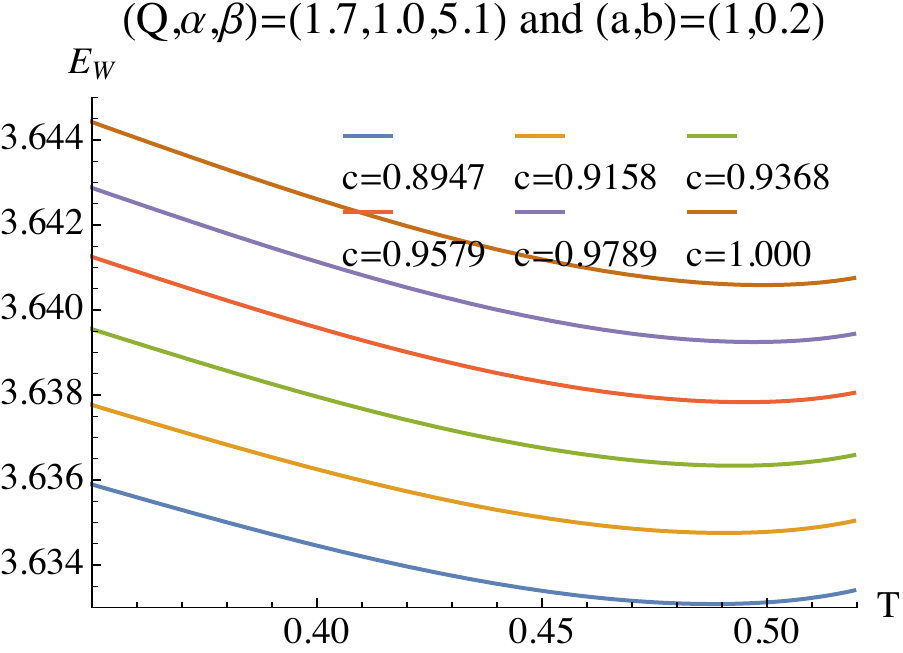}
  \caption{The temperature vs EWCS at the case where no thermal phase transition occurs.}
  \label{fig:eopnopt1}
\end{figure}

At last, we  {study} the case where no thermal phase transition occurs \red{at} $Q>b/3$. As we can see from Fig. \ref{fig:eopnopt1}, the EWCS again shows the opposite behavior from the MI (Fig. \ref{fig:noptmi1}) for relatively large temperature. For small temperatures, however, we can see that the EWCS starts to show similar behavior as that of the MI.

We summarize the behavior of EWCS in the massive gravity theory. First of all, EWCS can indeed reflect the first-order and second-order phase transitions of the system, which are represented by the jump and singular behavior at the critical temperatures, respectively. In addition, EWCS exhibits an opposite temperature dependence to MI. This shows that EWCS captures distinct degrees of freedom from that of MI. The behavior of MI can be understood by its association with HEE, whereas EWCS cannot. MI can still be dominated by the behavior of HEE in many cases, while EWCS is  {not controlled by} HEE.

\section{The critical behavior of the geometry-related quantities}\label{sec:critical}

From the above studies on HEE, MI and EWCS, it can be found that these geometry-related entanglement quantities have singular behaviors near the critical point of the second-order phase transition. In fact, we can analytically prove that these three quantities, even any geometrically related physical quantity, will exhibit the same scaling behavior. Moreover, we will also verify the correctness of the analytical analysis  {with numerical results}.

 {At the onset of the second-order phase transition,} any geometrically related physical quantity $A$ can be expressed as
\begin{equation}\label{eq:dadg}
  A = A_c + A' \delta g_{\mu\nu},
\end{equation}
 {where $A_c$ is the value at the critical point.}
Given that the entropy density behavior \eqref{eq:exponents}, we will find that,
\begin{equation}\label{eq:dadt}
   {\delta s}\sim \delta g_{\mu\nu} \sim (T-T_c)^{1/3}.
\end{equation}
Therefore, any geometry-related physical quantities will  {scale as,}
\begin{equation}\label{eq:scalingsnew}
  (A-A_c) \sim (T-T_c)^{1/3}.
\end{equation}

\begin{figure}
  \centering
  \includegraphics[width=0.48\textwidth]{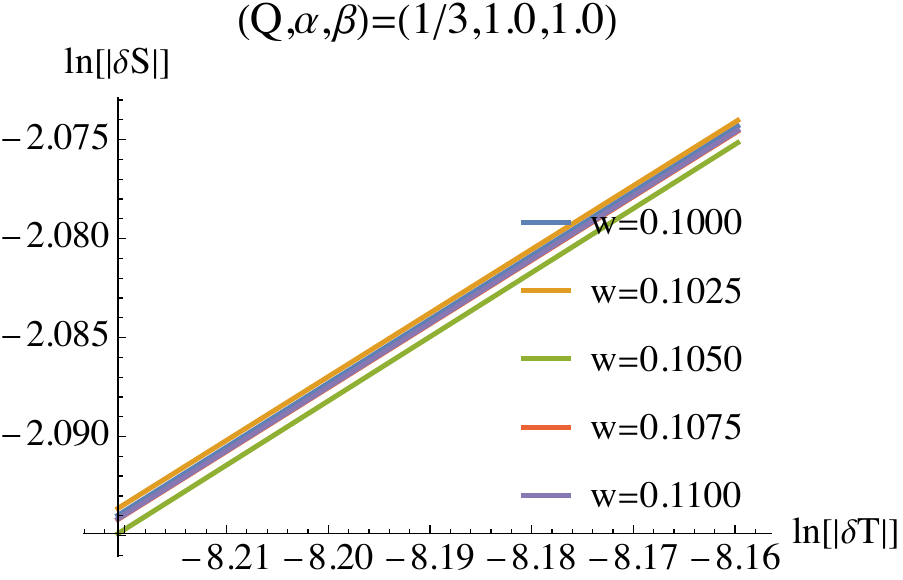}
  \includegraphics[width=0.48\textwidth]{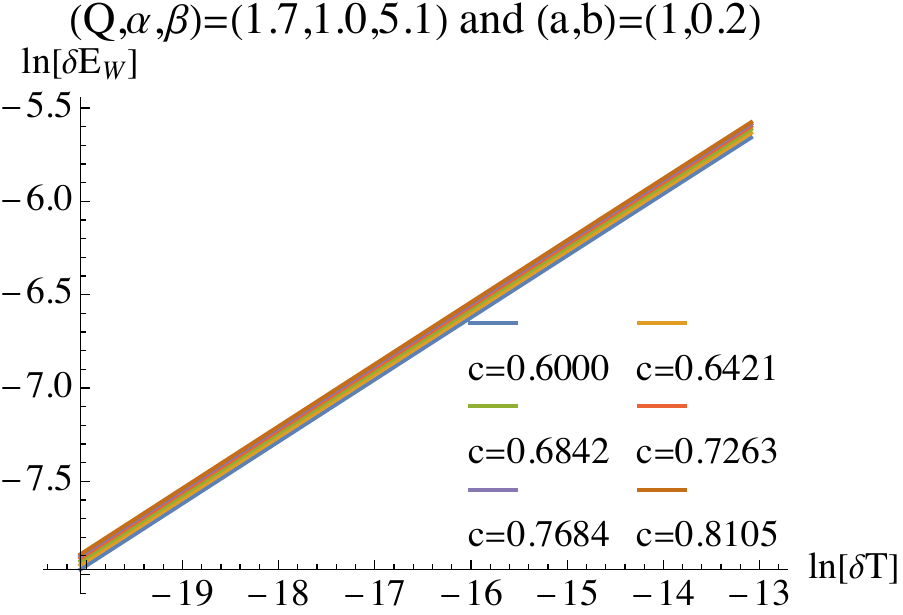}
  \caption{The scaling behavior between HEE and T, and EWCS and T.}
  \label{fig:scalingsnums}
\end{figure}

We support this analysis numerically. Fig. \ref{fig:scalingsnums} shows the scaling behavior of HEE with $T$ and EWCS with $T$ during the second-order phase transition. It can be seen from the figure that  {the} logarithm values  {of} $\delta S \equiv S-S_c$  {and} $\delta E_W \equiv E_W-E_{W_c}$ {depends linearly on the logarithm values of temperature difference $\delta T \equiv T-T_c$}, and the slopes converge to $1 / 3$.  {This means} that our analytical analysis and numerical analysis are mutually confirmed.

It is worth mentioning that in another second-order thermodynamic phase transition - holographic superconductor model, we also found a  critical behavior  {of the geometrically related quantities} \cite{Liu:2020blk}. Although {the scaling behavior exists in both the superconductivity phase transition and the Hawking-Page transition}, the underlying mechanisms are totally different. The core of all the critical behaviors in this paper comes from the \eqref{eq:dadg}, that is, the relationship between the change of metric  {$\delta g_{\mu\nu}$} near the critical point  {of the second-order Hawking-Page transition.} and the change of temperature  {$\delta T$}.
Notice that the Hawking-Page transition is not a traditional thermodynamic phase transition. For example, there is no spontaneous symmetry breaking and spontaneous condensation of order parameters. However, the holographic superconducting phase transition process is consistent with the traditional phase transition {theory}. When crossing the critical point, the $U(1)$ symmetry of the system is  {spontaneously} broken, accompanied by the spontaneous condensation of a complex scalar field.  {It} is the appearance of spontaneous condensation that modifies the background geometry. The back-reaction of the background geometry is the square of condensation correction \cite{Brito:2015oca}, which leads to the critical exponent as $\alpha =1$.  {While for} the thermodynamic phase transition in this paper, all geometry-related physical quantities have a critical exponent of $\alpha = 1 / 3$.

\section{Discussion}\label{sec:discuss}

In this paper, we studied the mixed state entanglement, as well as the entanglement entropy in the van der Waals phase transition of massive gravity. We find MI, EWCS, as well as HEE can all characterize the phase transitions. Also, the patterns of characterizing the phase transition for HEE, MI and EWCS are essentially the same: they encounter a jump at the first-order phase transition point and  {exhibit} a singular behavior at the second-order phase transition point. Moreover, HEE shows obvious configuration dependence. The HEE behavior with temperature is completely different in small configuration and large configuration.
 {As a comparison,}  {the temperature behaviors} of MI and EWCS {are} independent of the configuration.
However, we also observe an intriguing phenomenon that MI behavior with temperature is completely the opposite of that of the EWCS  in the critical region. Specifically, MI increases with decreasing temperature, while EWCS decreases with decreasing temperature. This phenomenon indicates that there is an important difference between EWCS and MI, especially in the critical regions. Moreover, we can see that MI and HEE are closely related in some cases, which means that EWCS might be a better candidate for the entanglement of mixed states. It is also proved that all geometry-related physical quantities have the same critical exponent $\alpha =  {1/3}$ near the second-order phase transition point, both analytically and numerically. Moreover, this phenomenon is completely different in the phenomenon and in the underlying mechanism from that of superconductivity phase transition.
 {The key reason is that the scaling relationships between the perturbation of the critical point metric perturbation (which affects the geometric quantities related to quantum information) and the temperature different are different. The superconductivity phase transition is accompanied by the appearance of condensation, but the Hawking-Page phase transition is not. This renders them have different scaling behaviors. This conclusion and related techniques can be applied to any physical quantity that only depends on geometric correlation.
}

In addition to the thermal phase transition accompanied by the emergence of order parameters \cite{Cai:2015cya} and the van der Waals-like phase transition, quantum phase transition may exhibit a completely different scenario. Quantum phase transition happens at absolute zero temperature when varying system parameters. In certain kinds of quantum phase transitions involving strong correlation, the order parameter does not emerge. It is then desirable to explore the characterization of these quantum phase transitions. In holographic duality theory, the quantum phase transition is related to the flow among different IR fixed points. However, solving a domain wall with IR fixed points is technically difficult since the zero-temperature limit may be accompanied by singular behaviors. Therefore, we are dealing with the quantum phase transitions at a low whereas nonvanishing temperature. Notice that the HEE, as well as MI in certain limits, can be determined by the thermal entropy, the HEE and MI are not good candidates for characterizing the quantum phase transitions. The EWCS, however, is by definition free of the control from the near horizon behavior. We can expect that the EWCS can work as a good diagnose of the quantum phase transitions.

\section*{Acknowledgments}

Peng Liu would like to thank Yun-Ha Zha for her kind encouragement during this work. This work is supported by the Natural Science Foundation of China under Grant No. 11575195, 11875053, 11805083, 11847055, 11905083, 12005077 and 11775036.

\end{document}